\documentclass[pra,aps,twocolumn,floatfix,superscriptaddress]{revtex4-1}

\usepackage {amssymb}
\usepackage {amsmath}
\usepackage{graphicx}
\usepackage{euscript}
\usepackage{txfonts}

\newcommand{\vect}[1]{{\mathrm {\mathbf #1}}} 
\newcommand{\pderiv}[2]{\frac{\partial #1}{\partial #2}} 
\newcommand{\reffig}[1]{Fig.~\ref{#1}}
\newcommand{\refeq}[1]{Eq.~(\ref{#1})}

\begin{document}


\title{Terahertz and higher-order Brunel harmonics: from tunnel to
  multiphoton  ionization regime in tailored fields}

\author{I. Babushkin}

\affiliation{Institute for Quantum Optics, Leibniz
  Universit\"at Hannover, Welfengarten 1, 30167 Hannover, Germany}

\affiliation{Max Born Institute, Max Born Str. 2a, 12489
    Berlin, Germany}

\author{C. Br\'ee}

  \affiliation{Weierstrass Institute, Mohrenstr. 39, 10117 Berlin, Germany}

\author{C. M. Dietrich}

\affiliation{Institute for Quantum Optics, Leibniz
  Universit\"at Hannover, Welfengarten 1, 30167 Hannover, Germany}

\author{A. Demircan}

\affiliation{Institute for Quantum Optics, Leibniz
  Universit\"at Hannover, Welfengarten 1, 30167 Hannover, Germany}

  \affiliation{Hannover Centre for Optical Technologies, Nienburgerstr. 17, 30167 Hannover, Germany}

\author{U. Morgner}

\affiliation{Institute for Quantum Optics, Leibniz
  Universit\"at Hannover, Welfengarten 1, 30167 Hannover, Germany}

  \affiliation{Hannover Centre for Optical Technologies, Nienburgerstr. 17, 30167 Hannover, Germany}

\author{A. Husakou}

\affiliation{Max Born Institute, Max Born Str. 2a, 12489
    Berlin, Germany}

\begin{abstract}
  Brunel radiation appears as a result of a two-step process of
  photo-ionization and subsequent acceleration of electron, without
  the need of electron recollision. We show that for generation of Brunel
  harmonics at all frequencies the subcycle ionization dynamics is of
  critical importance. Namely, such harmonics disappear at low
  pump intensities when the ionization dynamics depends only on the
  slow envelope (so called multiphoton ionization regime) and not on
  the instantaneous field. Nevertheless, if the pump pulse contains
  incommensurate frequencies, Brunel mechanism does generate new
  frequencies even in the multiphoton ionization regime.
\end{abstract}


\maketitle

\section{Introduction}

The study of the dynamics of ionization process attracted significant
attention last years, because laser sources allowing intense ultra-short
pulses with the control over the fast optical waveform shape became
available. Such studies gained importance also because 
generation of shortest possible pulses is based heavily on high
harmonic generation, which exploits the ionization dynamics.
Although the ionization process in most simple single-electron systems
such as noble gases is studied for several decades, its details are
still far from being fully clarified, thus being in a focus of recent
studies
\cite{yudin01a,blaga09,krausz09a,torlina15,shvetsov-shilovski16,ni16,pisanty16,jooya16,shafir13,wu12,paulus95}.
One of the most known peculiarities of the ionization process is its
strongly nonlinear dependence on the input field strength. For high
enough intensities ionization has predominantly a character of an
instantaneous tunneling, whereas for lower intensities it is
better described as rather slow multiphoton process. The boundary is
not exact and was several time revisited \cite{reiss08,topcu12}, but,
roughly speaking, it depends on the ratio of the ionization time
$\tau$ and the optical oscillation period $T$ known as Keldysh
parameter $\gamma=2\pi\tau/T$. The ionization process above this
threshold $\gamma\ll1$ is nearly instantaneous and depends on the
current field strength. This leads to an essentially step-wise
dynamics of ionization rate, with steps localized near extrema of the
electric field. On the other hand, below this threshold $\gamma\gg1$,
ionization depends more on the cycle-averaged field intensity rather
than on the instantaneous field. It should be noted also that for
multicolor fields the Keldysh parameters can not be defined anymore in
a universal way.

Creation of free electrons due to ionization leads to modification of
the material's refractive index, thus making the later field-dependent
and therefore introducing nonlinearity. The presence of nonlinearity
leads to generation of harmonics of the pump field, which are
sometimes called Brunel harmonics \cite{brunel90,balciunas13}.  We
remark that this mechanism does not requires return of the electron
back to the core and therefore involves the harmonics of relatively
low order (in contrast to high harmonics generation). In particular,
if the pump field is also asymmetric in time, this may
create also a net macroscopic asymmetric current which persists after
the pulse and thus gives rise to a low-frequency harmonic. This
process is a type of plasma-induced rectification and leads
to generation of frequencies in terahertz (THz) range
\cite{cook00,kress04,bartel05,reimann07,kim08b,babushkin10a,babushkin11},
or the ``zero-order'' Brunel harmonic. Higher order harmonics are also
generated, for both single- and multi-color fields, although with
decreasing efficiency.  Brunel harmonics are to be distinguished from
the ones appearing due to dynamics of the bounded electrons (so called
four-wave-mixing process). In general, experimental differentiation
between the two is a rather complicated task
\cite{bree11a,wahlstrand11, brown12,bejot10,serebryannikov14}.

In this article we show that, looking from the perspective of the
classical electron motion in the continuum, the Brunel harmonics
appear to be extremely sensitive to the subcycle step-wise character
of the ionization process. If the ionization is ``fast enough''
(tunnel ionization regime), it gives rise to the new harmonics whereas
if the ionization ``slow'' (multiphoton ionization regime), Brunel
harmonics do not appear. We show the transition between these two
regimes for the pump pulses containing only one frequency. We also
prove that this is the case for the THz harmonics in two-color
case. Furthermore, we show that the situation may change if the
two-color field contains incommensurate harmonics, that is harmonics
with the non-integer frequency ratio. Such harmonics lead to beatings
in the envelope which can be slow enough to switch on the Brunel
mechanism again, even if the ionization process is slow.

The article is organized as follows: in Sec.~\ref{sec:el-traj} we make a short introduction into the 
mechanism of  Brunel harmonics generation and
describe the details of ionization process relevant to our consideration. In
Sec.~\ref{sec:single-col} we analyze the Brunel harmonics in tunnel
and multiphoton ionization regime in single-color pump whereas
Sec.~\ref{sec:2-col-comm} and Sec.~\ref{sec:2-col-incomm} are devoted
to two-color pulses, with commensurate and incommensurate frequencies,
respectively. The analytical description is developed and discussed in
Sec.6 
followed by concluding remarks 
Sec.~\ref{sec:concl}.

\section{\label{sec:el-traj} Electron trajectories, macroscopic
  plasma current, and ionization}

Brunel radiation can be described classically as a radiation created
by a current of all free electrons. In particular, assuming a small
volume of gas (smaller than the considered wavelengths) ionized in a
pump field $E(t)$, the Brunel radiation created by the free electron
current $J(t)$ in this volume will be:
\begin{equation}
  \label{eq:4}
  E_\mathrm{Br}(t)=g\frac{dJ(t)}{dt},
\end{equation}
where $g$ is a factor depending on the observation point. If delays in
light propagation can be neglected but still the distance to the
observer $r$ is large enough, is given as \cite{babushkin11}:
$g=\delta V/(4\pi\varepsilon_0 c^2 r)$, where $\delta V$ is the volume
of the plasma spot. In the current article we assume argon as a
working gas.

The expression for the current is given by $J(t)=-e\rho(t) v(t)$,
where $e$, $\rho(t)$, $v(t)$ are the electron charge, density of
electrons and their average velocity, respectively.
The change of current $\delta J$ after time moment
$\delta t$ can be expressed as
$\delta J=-e(\rho(t) \delta v+ v(t)\delta\rho)$. When calculating the
change of average velocity $\delta v$, we take into account that the newly
ionized electrons have zero average velocity, thus reducing $v$:
$\delta v = -eE(t)\delta t/m_e-\delta\rho v(t)/\rho(t)$, where $m_e$
is the electron mass. As a
result, we get a remarkably simple expression
\begin{equation}
  \label{eq:simple}
  E_\mathrm{Br}(t)=g\frac{dJ(t)}{dt}=\frac{ge^2}{m_e}E(t)\rho(t).
\end{equation}

The electron density $\rho(t)$ is related to the
ionization rate $W(t)$ as
\begin{equation}
\pderiv{\rho(t)}{t} = (\rho_0-\rho(t))W(t), \label{eq:rho}
\end{equation}
 where $\rho_0$ is the density of non-ionized atoms before interaction
 with the pulse.
 Ionization rate dynamics is important and is described below.

\begin{figure*}
\centering 
\includegraphics[width=0.8\textwidth]{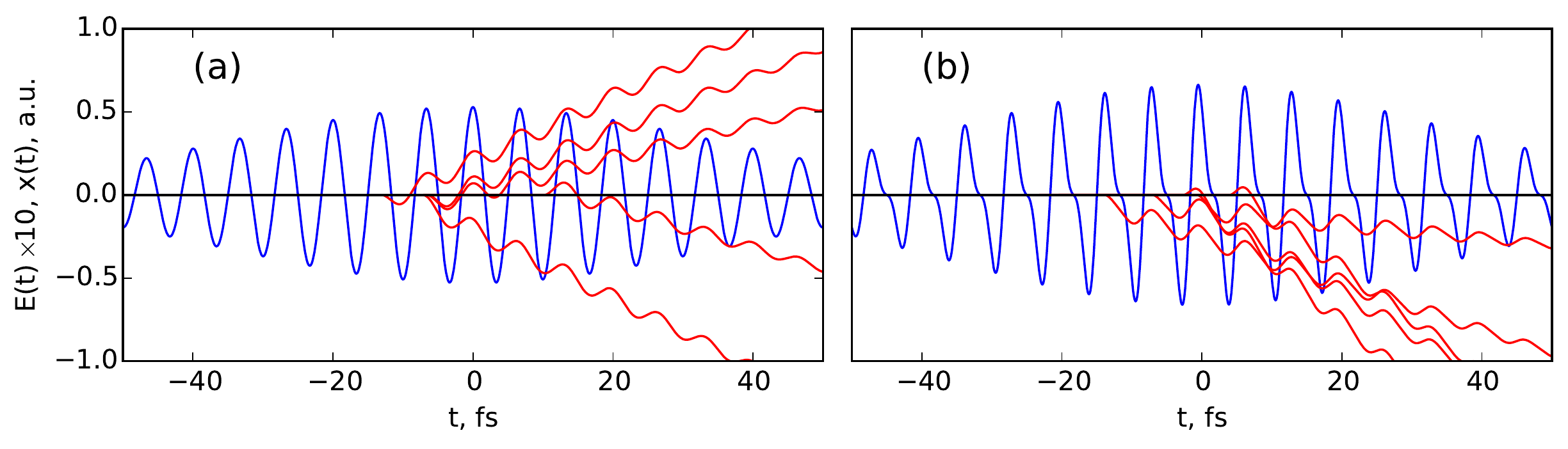}
\caption{Trajectories $x(t)$ (red lines) of sample electrons ionized
  in the electric field $E(t)$ (blue line) of the strong one-color (a)
  and two-color (b) optical pulses (parameters are given in the text).
  In (b), because of presence of two colors, the field shape is
  slightly asymmetric which leads to an asymmetry between up- and
  down- directions in the electron trajectories and thus to
  significant change of the resulting macroscopic current $J(t)$ and
  thus the Brunel radiation $E_{Br}(t)$. This
  illustrates the importance of subcycle dynamics of the ionization
  and movement.} \label{fig:trajectories}
\end{figure*}

 In \reffig{fig:trajectories} one can see an exemplary process of
 free electron creation and their dynamics for single- and two-color pump field.
We consider the electric field $E(t)$ in the form: 
\begin{equation}
  \label{eq:5afirst}
  E(t) =  F_1(t) \cos{\omega_0t} + F_2(t) \cos{(2\omega_0t+ \phi_{12}(t))},
\end{equation}
where $\phi_{12}(t)$ is the phase which we allow to vary slowly in time,
and 
\begin{equation}
  \label{eq:1}
  F_i(t) = \sqrt{2I_i(t)/(c\epsilon_0)},\,\,
  I_i(t)=I_{i0}e^{-t^2/T_i^2},\, F_{0i}=\sqrt{2I_{0i}/(c\epsilon_0)}.
\end{equation}
are slowly varying field amplitudes and intensities which describe the
corresponding pulse envelope, $T_i$ is the pulse duration, and
$\phi_{12}$ is the relative phase.

In \reffig{fig:trajectories}, several exemplary trajectories have been
generated with the electron ionized randomly with the probability
corresponding to the tunnel ionization rate as discussed below, for
the cases of $I_2=0$ in \reffig{fig:trajectories}(a) and $I_2=20$
TW/cm$^2$ in \reffig{fig:trajectories}(b). In both cases we assumed
$I_1=100$ TW/cm$^2$, the frequency $\omega_0$ corresponding to 2
$\mu$m wavelength, and pulse duration of 50 fs. For the parameters
chosen here, the electrons are ionized with highest probability close
to the maxima of electric field and by subsequent acceleration move,
in average, away from the origin. Neither recollision no core
potential were taken into account in our simulations.  Remarkably, the
net current in the case of a two-color field in
\reffig{fig:trajectories}(b) possesses a spatial asymmetry, which is
absent in the single-color case. Therefore, the net current $J(t)$ is
also very different in both cases. Already this simple example shows
that the net current $J(t)$ and thus the Brunel radiation $E_{Br}$ can
be significantly modified by the field dynamics on the subcycle scale.

Here we will show that the dynamics of ionization process itself can be
also of critical importance for the macroscopic current buildup. 
As far as we know, there is no universally acceptable model which describes
ionization dynamics for all intensities on all relevant timescales
which is also valid for multicolor field. As it was mentioned before,
for high intensities ($\gamma\ll1$) the ionization takes place on the
subcycle scale. In this case, the ionization rate depending on the
instantaneous field strength, describing tunneling process in
quasi-static approximation, is acceptable for arbitrary pulse
shapes and thus arbitrary number of frequencies in the pulse.
We use here the static tunneling approximation formula in the
form \cite{ammosov86a}:   
\begin{equation}
  \label{eq:11}
  W_{ST}(t) = \frac{\alpha}{\tilde E(t)} \exp\left\{-\frac{\beta}{\tilde E(t)}\right\},
\end{equation}
where $W_{ST}(t)$ is the ionization rate, $\tilde E(t) = |E(t)|/E_a$,
$E_a \approx 5.14 \times 10^{11}$ Vm$^{-1}$ is the field unit in
atomic system of units (a.u.), coefficients $\alpha = 4
\omega_a r_H^{5/2}$ and $\beta = (2/3) r_H^{3/2}$ are defined through
the ratio of ionization potentials of argon and hydrogen atoms
$r_H=U_{Ar}/U_H$, $U_H=13.6$ eV, $U_{Ar}=15.6$ eV, and $\omega_a
\approx 4.13 \times 10^{16}$ s$^{-1}$.

For lower intensities $\gamma\gg1$, when ionization probability is
much less then an optical cycle, ionization rate depends more on the
cycle-averaged intensity - that is, on the field envelope rather than
instantaneous field strength. The tunnel ionization formula
\refeq{eq:11} does not catch this dynamics anymore.  To calculate
corresponding ionization rate for both single- and two-color fields we
used the generalization of the well known Keldysh-Faisal-Reis theory
\cite{keldysh65,faisal73,reiss80} for the field with arbitrary number
of harmonics, which was developed by Guo, Aberg and Crasemann
\cite{guo92} and which we abbreviate here as GAC model. In
App.~\ref{sec:multiphot} the corresponding ionization rate for the
case where the field contains two colors $\omega_0$, $2\omega_0$.
This model provides us the dependence of the ionization rate on both
intensities $I_1(t), I_2(t)$ and the relative phase $\phi_{12}$ in
explicit (though cumbersome) formalism:
\begin{equation}
W_{GAC}(t)=W_{GAC}(I_1(t),I_2(t),\phi_{12}),\label{eq:2}
\end{equation}
where $W_{GAC}$ is given by  \refeq{eq:W} 
and
described in details in Appendix~\ref{sec:multiphot}.
Importantly, now ionization does not depend on the instantaneous field
but rather on the \textit{slow envelopes} of the harmonics as well as on their
phase $\phi_{12}$.

Finally, for the case of a \textit{single-color} field there is a 
relatively simple formula derived by Ivanov and Yudin \cite{yudin01a},
which correctly describes both cycle-averaged ionization rate and its
slow and subcycle dynamics in both tunnel and multiphoton regimes. We
will refer to this model as IY one, and the corresponding ionization
rate is given by:
\begin{equation}
  \label{eq:3}
  W_{IY}(t) = \alpha \exp{\left(-\beta'F(t)^2-\beta''E(t)^2\right)}, 
\end{equation}
where the field is assumed to be in the form \refeq{eq:5afirst} with
$F_2=0$, thus $F(t)\equiv F_1(t)/E_a$  is the slow field envelope in
\refeq{eq:1} in the units of $E_a$, and
the coefficients are defined as:
\begin{gather}
  \label{eq:5}
  \alpha = \frac{\pi C_k^2
    }{\tau}\left(\frac{2\kappa^3}{F}\right)^{2Q/\kappa},\,
    \beta' = \frac{2 \sigma_0 + \sigma_0''}{2\omega_0^3},
    \beta''=\frac{\sigma_0''}{2\omega_0^3}, \\
  \sigma_0=\frac12 (G^2-\frac12)\ln{C}-G\frac{\gamma}{2},\,
     \label{eq:5b} \\
  \sigma_0''=\ln{C}-2\frac{\gamma}{G}, \, G=\sqrt{1+\gamma^2},\, C = 2G^2-1 + 2G\gamma.
    \label{eq:5bb}
\end{gather}
Here $\kappa=\sqrt{2r_H}$ ($r_H$ is defined below \refeq{eq:11}),
$\tau=\kappa/F$ is the tunneling time, $\gamma=\omega_0\tau$ is the
Keldysh parameter, $Q$ is the effective charge of the ionic core
($Q=1$ for noble gases), $C_k$ is related to the radial asysmptote of
the quantum orbital and in our case also $C_k\approx 1$.
This formula  is not applicable for multicolor
fields, but for single-color ones it provides a possibility to
inspect 
the transition between tunnel and multiphoton regime in details, which
is done in the next section.

\section{\label{sec:single-col} Brunel harmonics in single-color fields}

\begin{figure*}
\centering 
\includegraphics[width=0.8\textwidth]{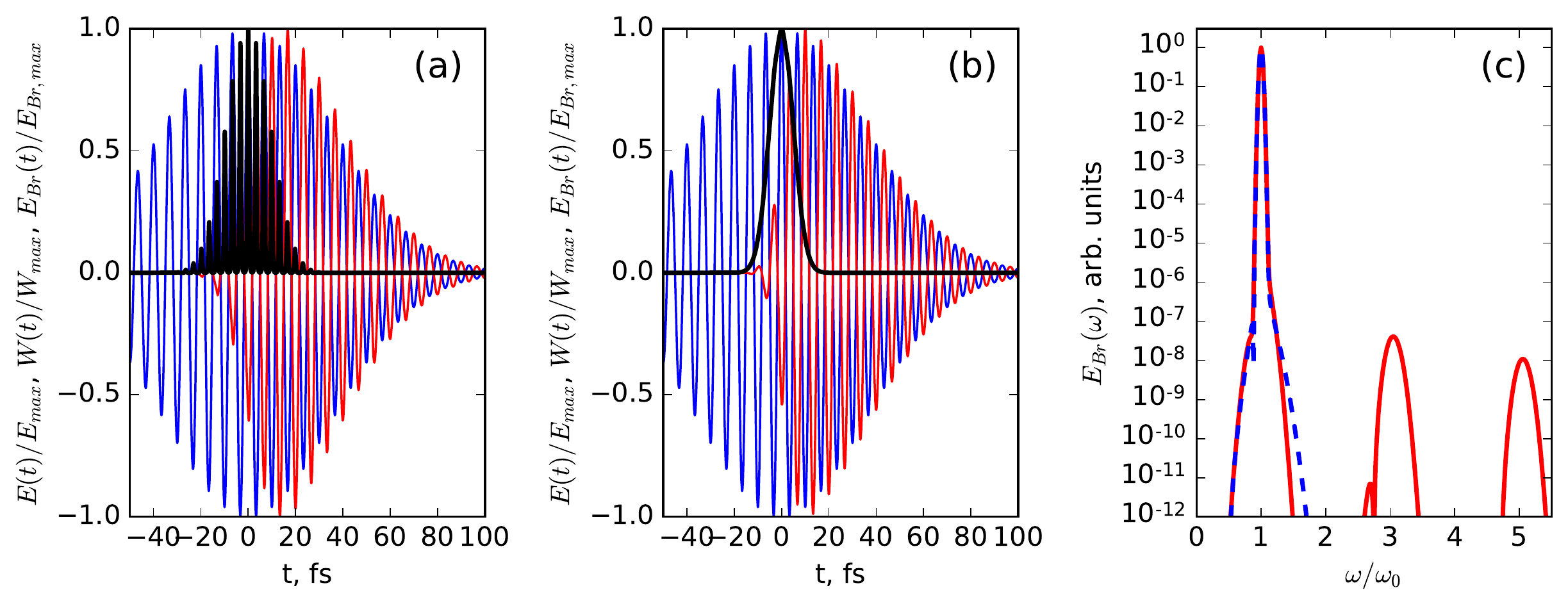}
\caption{Brunel radiation in the tunnel and multiphoton regimes in the
  case of single color pulse. In (a) and (b), the field (blue line),
  ionization (black line) and $E_{Br}(t)$ (red line) are shown for
  $F_{01}=0.053$ a.u. (a) and $F_{01}=2.6\times10^{-4}$ (200 times
  less) (b). The other parameters as in \reffig{fig:trajectories}(a).
  In (c), the spectra of Brunel radiation $E_{Br}(\omega)$
  for (a)  (red line) and (b) (blue
  dashed line) are shown.
} \label{fig:1-col}
\end{figure*}

We start our consideration of Brunel radiation from discussing the
single-color field ($I_{02}=0$) in \refeq{eq:1},~\refeq{eq:5afirst}.
We use \refeq{eq:simple} together with equation for the free electron
density \refeq{eq:rho} and ionization calculated using IY formula
$W=W_{IY}$ \refeq{eq:3}.  In \reffig{fig:1-col} the dynamics of the
electric field and ionization as well as the Brunel radiation is shown for the
parameters as in \reffig{fig:trajectories}(a) but with two different
intensities corresponding to the tunnel [\reffig{fig:1-col}(a)] and
multiphoton [\reffig{fig:1-col}(b)] regime. Of course, the bare
ionization rate $W$ in the case of the weak field is many orders of
magnitude lower than in the tunnel one. We are interested in the
dynamics, thus the ionization rate as well as Brunel radiation
$E_{Br}(t)$ is renormalized to their maximum values in every case. The
ionization rate in the tunnel regime demonstrates strong
subcycle-scale dynamics whereas in the multiphoton regime it changes
on much longer timescale of the slow field envelope.  The Brunel
radiation $E_{Br}(t)$ looks not very different in the time domain, but
significant difference can be observed in the frequency range shown in
\reffig{fig:1-col}(c). Namely, one can see that the Brunel harmonics
at the frequencies $\omega=n\omega_0$, $n=3,5,7\ldots$ indeed arise if
the ionization is in the tunnel regime (red lines) and disappears if
multiphoton ionization takes place (blue lines). As it is shown in the next section, this is
because the field of Brunel harmonics appears as a convolution of the
fast optical field with the electron density, and this convolution has
no new harmonics if the $\rho(t)$ changes slowly in time, as it is the
case for the multiphoton regime. Note that the
harmonic at $\omega=\omega_0$ is related related to the free electron
movement (in contrast to electron birth) and is not interesting for us
in the present article.  

\begin{figure}
\centering 
\includegraphics[width=\columnwidth]{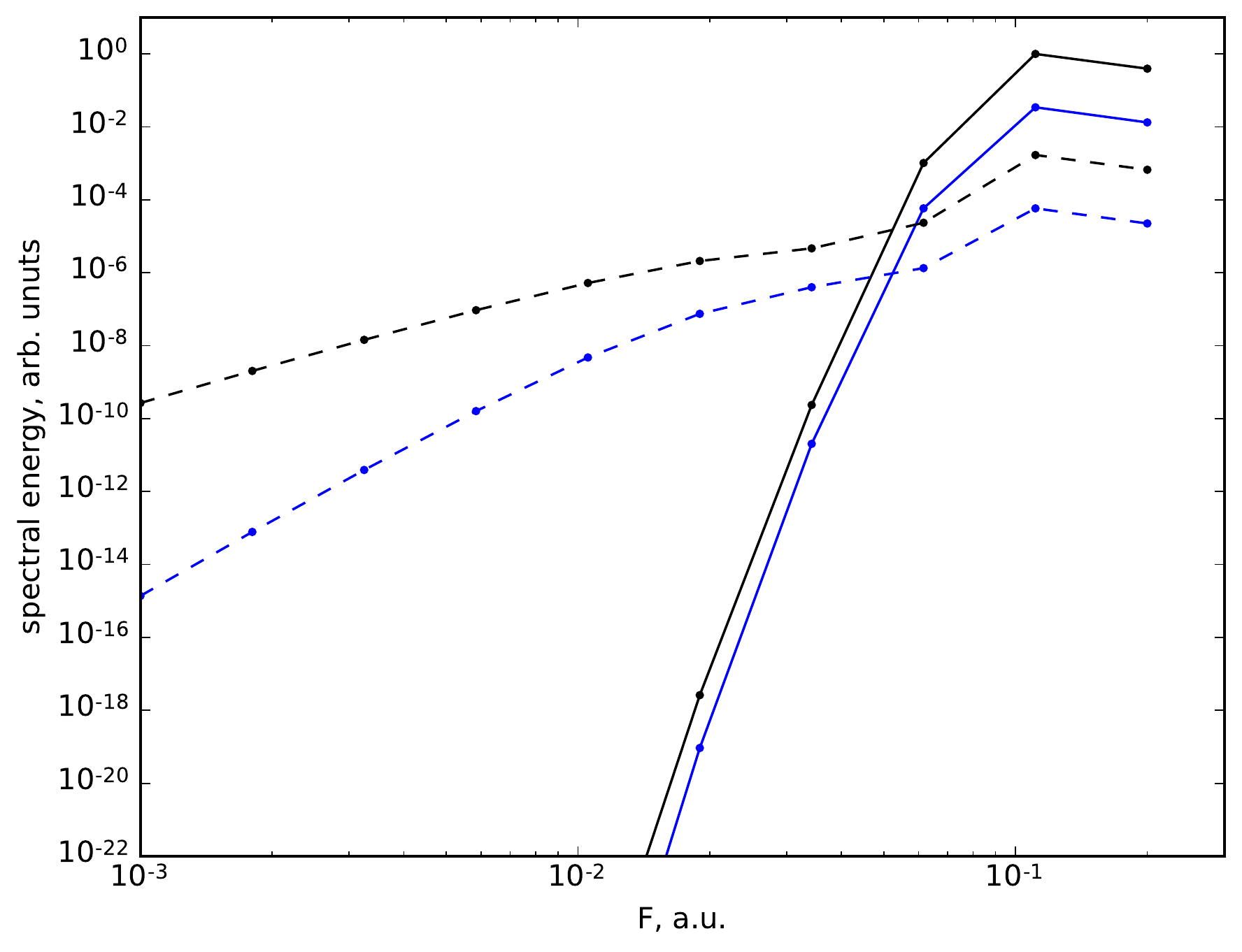}
\caption{The energy of Brunel harmonics in dependence on the peak
  amplitude $F_{01}$ of a single-color pulse with the parameters as in
  \reffig{fig:trajectories}(a). Black (blue) lines show
  the energy of the 3th (5th) harmonics. The absolute energies are
  shown in solid lines whereas the dashed lines represent the energies
  per single ionized atom. } \label{fig:par}
\end{figure}

Using IY formula it is possible to observe the transition
between the two types of dynamics. Such transition is shown in
\reffig{fig:par}, where the energies of the 3th and 5th Brunel harmonics
are shown (solid black and blue lines, respectively) in dependence on
the peak amplitude $F_{01}$ of the single-color field with the 
parameters as in \reffig{fig:trajectories}(a). As the field
amplitude decreases, the intensities of the Brunel harmonics decrease
rather dramatically. There are two effects in this, one is
connected to the overall density of free electrons, which also decreases,
the other is connected to the ionization dynamics on its own. To
demonstrate clearly the later effect, we plotted the energies  rescaled
per single ionized atom (technically, to obtain such quantities we
rescale the ionization current $J$ and thus $E_{Br}$ to
$\rho(t=\infty)/\rho_0$). The resulting quantities are shown in
\reffig{fig:par} by dashed lines.

If the Brunel harmonic energies depended only on the number of
ionized electrons and not on the corresponding (subcycle) ionization
dynamics, such rescaled energies would not depend on $F$. However,
this is not the case.As one can see, they also decay quickly 
with decreasing $F$ (although with much lower rate), demonstrating
the importance of the fast tunnel ionization dynamics. This decay is
also much faster for the 5th harmonics than for the 3th one, indicating
strong nonlinearity of the process.  As it will be shown in
Sec.~6,
 the appearance of 3th and 5th harmonics would be 
impossible if the ionization rate were truly slow (contains no harmonics at
the pump frequency and higher). This shows also that the fast
tunneling component in ionization dynamics is still present also for
very low fields.  

\section{\label{sec:2-col-comm} Brunel harmonics in two-color fields:
  commensurate case}

\begin{figure*}
\centering 
\includegraphics[width=0.8\textwidth]{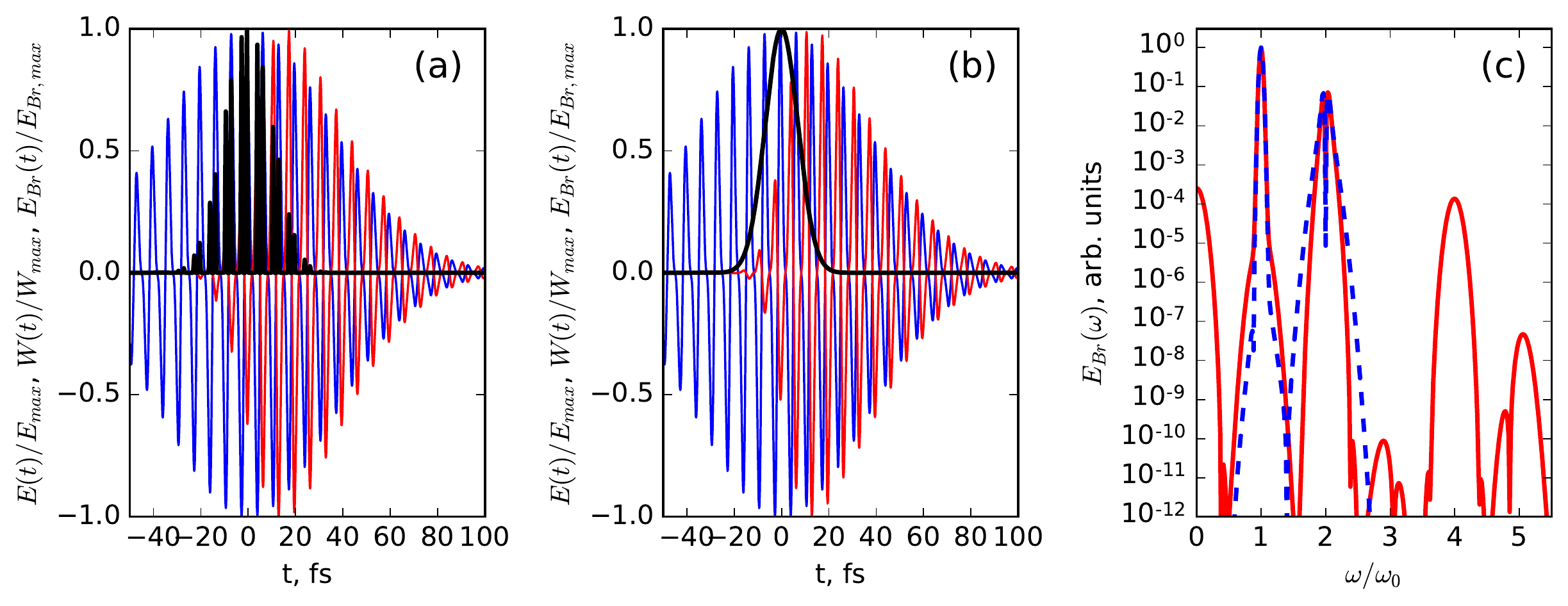}
\caption{Brunel radiation in the tunnel and multiphoton regimes in the
  case of two-color pulse. In (a) and (b), the field (blue line),
  ionization (black line) and $E_{Br}(t)$ (red line) are shown for
  $F_{01}=0.053$ a.u. (a) and $F_{01}=2.6\times10^{-4}$ a.u. (200 times
  less) (b). The other parameters as in \reffig{fig:trajectories}(b). In (c), the spectra of Brunel radiation $E_{Br}(\omega)$
  for (a) (red line) and (b) (blue
  dashed line) are shown.     
} \label{fig:2-col}
\end{figure*}

Now let us return to the case when the pump pulse consists of the
fundamental frequency $\omega_0$ and its second harmonics $2\omega_0$,
that is, \refeq{eq:5afirst},~\refeq{eq:1} with $I_{02}\ne0$,
$\phi_{12}=\mathrm{const}$.  In this case, both odd and even harmonics
can appear. That is, Brunel harmonics are located near frequencies
$\omega=n\omega_0$ with $n=0,2,3,4\ldots$. In particular, a zero one,
which is located typically in realistic setups at THz frequency, is of
special importance because THz radiation is difficult to generate.
The appearance of even harmonics is related to the asymmetry of the
field as shown in \reffig{fig:trajectories}(b).  For the two color
case, as it was mentioned before, there is no general formula for
ionization process allowing to track the transition from tunnel to
multiphoton regime as we did this in \reffig{fig:par}. Nevertheless,
we can consider these two cases separately, using tunnel ionization
rate $W_{ST}$ (\refeq{eq:11}) for higher intensities and GAC
ionization rate $W_{GAC}$ for lower ones as discussed in
Sec.~\ref{sec:el-traj}.

The corresponding figure for two-color case is presented in
\reffig{fig:2-col}. The parameters are the same as in
the single-color case in \reffig{fig:1-col}, but now $I_{02}\ne0$,
namely $I_{02}/I_{01}=0.2$ as in \reffig{fig:trajectories}(b). In
particular, we have taken $\phi_{12}=\pi/2$ which delivers the maximal
energy for the THz harmonic \cite{kim07}. 
One can
see that the same behavior is observed: in the tunnel ionization
regime the Brunel harmonics at $\omega=n\omega_0$ are visible whereas
in the multiphoton regime they disappear. This behavior is completely
analogous to the single-color case and takes place because in the
multiphoton regime ionization is ``too slow'' to create harmonics as
discussed further in Sec.~6. 

\section{\label{sec:2-col-incomm} Brunel harmonics in two-color fields:
  incommensurate case}

Let us now try two-color pump fields with incommensurate frequencies.
By incommensurate frequencies we assume the ones with
the ratio between them being 
not-integer. For this, we slightly detune 
the second harmonic $2\omega_0$
 to $2\omega_0+\Delta\omega$ with $\Delta\omega\ll
\omega_0$. That is,
the electric field we consider  is: 
\begin{equation}
  \label{eq:5a}
  E(t) =  \sqrt{2I_1(t)/(c\epsilon_0)} \cos{\omega_0t} + \sqrt{2I_2(t)/(c\epsilon_0)}
  \cos{(2\omega_0t+\Delta\omega t + \phi_{12})}.
\end{equation}
The idea behind consideration of such field shape is the following:
slight detuning $\Delta\omega\ll \omega_0$ leads in some slow beatings
in the field amplitude, and in this case even ``slow'' ionization in
multiphoton regime may ``see'' these beatings and thus generate
some harmonics. 

\begin{figure*}
\centering 
\includegraphics[width=0.8\textwidth]{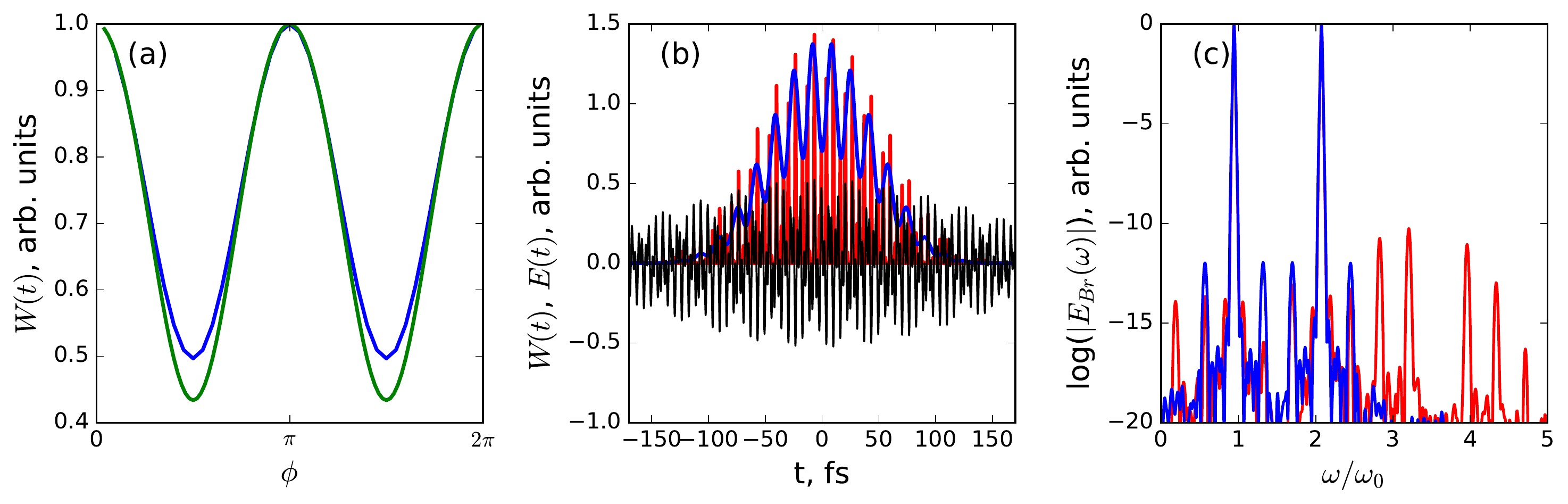}
\caption{ Brunel harmonics for incommensurate frequencies. (a) Two
  color multiphoton ionization rate $W_{GAC}(I_1,I_2,\phi)$
  (normalized to the maximum over all $\phi$) depending on phases
  between two harmonics $\phi$ for $I_1=I_2=20$ TW/cm$^2$ (blue line)
  and $I_1=I_2=10$ TW/cm$^2$ (green line). (b) The field (black line),
  tunnel ionization rate (red line) and multiphoton ionization rate
  (blue line) for a 200 fs pulse containing frequencies $\omega_0$ and
  $2.2\omega_0$ with $I_1=I_2=20$ TW/cm$^2$. (c) Brunel harmonics (per
  single ionized electron) due to tunnel ionization (red line) and
  multiphoton ionization (blue line). } \label{fig:nonharm}
\end{figure*}

For the high intensity in the tunnel ionization regime we can still
use the tunnel formula \refeq{eq:11}. On the other hand, the GAC
formalism is valid for \textit{commensurate} frequencies.  However, we
note that with the condition $\Delta\omega\ll \omega_0$, we can
interpret the frequency shift as additional relative phase which is
slowly varying in time:
\begin{equation}
  \phi_{12}(t) \to \phi_{12} + \Delta\omega t
  \label{eq:6}
\end{equation}
and thus consider the field in the form \refeq{eq:5afirst} with
$\phi_{12}(t)$ given by \refeq{eq:6}.  We then use the GAC formalism
for two-color pump with frequencies $\omega_0$, $2\omega_0$ and the
corresponding ionization rate
\begin{equation}
  \label{newW}
W(t)=W_{GAC}(I_1(t),I_2(t),\phi_{12}+ \Delta\omega t),
\end{equation}
with $W_{GAC}$ given in \refeq{eq:W}. 

As one can see in Fig. \ref{fig:nonharm}(a), where the dependence of
the multiphoton ionization rate on the phase between pulses for
exemplary intensities $I_1$, $I_2$ is shown, the change of the
effective phase effective phase $\Delta\omega t + \phi_{12}$ lead to
noticeable modification of multiphoton ionization rate. Indeed, as one
can see in the Fig. \ref{fig:nonharm}(b), not only tunnel ionization
rate but also multiphoton one varies across the pulse.  For this case
we used longer pulses ($T_{1,2}=200$ fs) to make slow beatings possible, with the intensities $I_1=I_2=100$ TW/cm$^2$ for the tunnel case
and $I_1=I_2=20$ TW/cm$^2$ for the multiphoton one.

The spectrum of the radiation generated by the Brunel mechanism for
$\Delta\omega=0.2\omega_0$ is presented in
\reffig{fig:nonharm}(c). One can see the generation of new frequencies
in the vicinity of the pump frequencies in the form of a comb-like
pattern with spacing of roughly $0.4\Delta\omega$ both in the case of
tunnel and multiphoton ionization regimes. This is in contrast to the
pump pulse which have peaks only at $\omega_0$,
$2\omega_0+3\Delta\omega$.  In the multiphoton regime, only few such
frequencies is generated [blue line in
\reffig{fig:nonharm}(c)]. Effectively, only frequencies
$\omega_0\pm\Delta\omega$, $2\omega_0+\Delta\omega\pm 3\Delta\omega$
appear in this case.  In contrast, in the tunnel ionization regime,
this comb contains much more frequencies.

\section{Analytical considerations of Brunel
  harmonics \label{sec:an11}}

To get a further insight into the processes governing the generation
of new frequencies by Brunel mechanism, we present the quantities
$E_{Br}(t)$, $E(t)$, and $\rho(t)$ as Fourier series near the maximum
of the pump pulse. This method was first proposed in \cite{brunel90} and
further developed in \cite{babushkin11,babushkin15,balciunas15}. Here we
neglect the slow amplitude changes due to Gaussian envelopes, that is,
we consider pulses formally infinite in time to keep consideration
simple. Under this condition, we have for the pump pulses with
commensurate frequencies:
\begin{gather}
  \label{eq:d1}
  \nonumber
  E_{Br}(t) = \sum_n E_{Br,n}e^{in\omega_0 t},\,
  E(t) = \sum_n E_ne^{in\omega_0 t},\, \\
  \rho(t) = \sum_{n} \rho_ne^{in\omega_0 t},
  W(t) = \sum_{n} W_ne^{in\omega_0 t}.
\end{gather}
If $\rho\ll\rho_0$ we can also assume $\partial_t\rho\approx\rho_0W$,
that is:
\begin{equation}
  \label{eq:10}
  \rho_n = \frac{\rho_0}{i\omega}W_n.
\end{equation}

By substituting the series \refeq{eq:d1} into Eq. (\ref{eq:simple}),
multiplicating the result by $e^{-in'\omega_0 t}$ and integrating over
time (assuming that
$\int_{-\infty}^{+\infty}e^{it\omega_0(n-n')}dt=2\pi\delta_{n,n'}$,
where $\delta_{n,n'}$ is the Kronecker delta),
we get:
\begin{equation}
E_{Br,n}=\frac{ge^2}{m_e}\sum_jE_{j}\rho_{n-j}.
\label{series}
\end{equation}

First, consider the situation arising when we are in multiphoton
ionization regime. In this case, $\rho$ contains only the component
$\rho_0$. That is,
\begin{equation}
  \label{eq:8}
  \rho(t) = \rho_0.
\end{equation}
In this case, the harmonics present
in $E_{Br,n}$ coincide with the ones in the pump $E_n$:
\begin{equation}
  \label{eq:9}
  E_{Br,n} = \frac{ge^2}{m_e}\rho_0 E_n,\, E_{Br}(t) =  \sum_n E_{Br,n}e^{in\omega_0 t}.
\end{equation}
One can see that the  slow dynamics 
of the electron density does not lead to generation of
new harmonics, but rather to modification of the existing ones. This
is valid for arbitrary pump spectra (with commensurate frequencies). 

In contrast, in the case of tunneling ionization, $\rho(t)$ increases
steeply each half optical period, and contains strong components with
$n\neq 0$. In this case, the
Brunel harmonics appear. In particular, if $E$ contains only a single
frequency (that is, only $E_1,E_{-1}\ne0$), then from \refeq{series} we have:
\begin{equation}
  \label{eq:12}
  E_{Br,n}=\frac{ge^2}{m_e}\left(E_1\sum_j\rho_{n-1} + E_{-1}\sum_j\rho_{n+1}\right).
\end{equation}
In this case, only even components $\rho_n$, $n=2N$ are present, which
results, according to Eq. (\ref{series}), in appearance of odd
harmonics in $E_{Br}(t)$. In contrast, for the two-color pump (we do
not write the explicit formula for this case here), all
components are present in $\rho_n$, and accordingly also in
$E_{Br,n}$, including those with $n=0$.

In the case of non-commensurate frequencies, the series (\ref{eq:d1})
 must be extended to include components with frequency
$\Delta\omega$ and its multiples:
\begin{gather}
  \label{eq:dd1}
  E_{Br}(t) = \sum_{n,m} E_{Br,n,m}e^{it(n\omega_0+m\Delta\omega)},\,
  E(t) = \sum_{n,m} E_{n,m}e^{it(n\omega_0+m\Delta\omega)},\,\\
  \rho(t) = \sum_{n,m} \rho_{n,m}e^{it(n\omega_0+m\Delta\omega)},
  W(t) = \sum_{n,m} W_{n,m}e^{it(n\omega_0+m\Delta\omega)}.
  \label{eq:dd3}
\end{gather}
Substituting into \refeq{eq:simple} and multiplying by $e^{-it(n''\omega_0+m''\Delta\omega)}$, we will have:
\begin{gather}
  \nonumber
  \sum_{n,m} E_{Br,n,m}e^{it(\omega_0(n-n'')+\Delta\omega(m-m''))}= \\ =
  \sum_{n,m,n',m'}
  E_{n,m}\rho_{n',m'}e^{it(\omega_0(n+n'-n'')+\Delta\omega(m+m'-m''))},
  \label{eq:long-series}
\end{gather}
We can now integrate over time as we did by derivation of
\refeq{series}. 
In the case if there is no such $N$, $M$ that
$N\omega_0+M\Delta\omega=0$ we can use the relation:
$\int_{-\infty}^{+\infty}e^{it(\omega_0(n-n'')+\Delta\omega(m-m''))}dt=2\pi\delta_{n,n'}\delta_{m,m''}$,
thus obtaining:
\begin{equation}
  \label{eq:13}
  E_{Br,n,m}=\frac{ge^2}{m_e}\sum_{j,i}E_{j,i}\rho_{n-j,m-i}.
\end{equation}
This equation is relatively good applicable in our situation 
since we need large $N$, $M$ to get $N\omega_0+M\Delta\omega=0$, which
would require large number of harmonics in the spectrum.

In the multiphoton regime, because multiphoton ionization is fast enough to
resolve beatings with frequency $\Delta\omega$ but still slow on the
scale of fundamental frequency $\omega_0$, we
have:
\begin{equation}
  \label{eq:14}
  \rho(t) = \sum_m \rho_{0,m} e^{im\Delta\omega t},
\end{equation}
which is to be compared to the multiphoton commensurate case
\refeq{eq:8}. Because of this, instead of \refeq{eq:9} we will have:
\begin{equation}
  \label{eq:15}
    E_{Br,n,m}(\omega) = \frac{ge^2}{m_e}\sum_j\rho_{0,j} E_{n,m-j}.
\end{equation}

As 
follows from 
the symmetry considerations and as one can see in
\reffig{fig:nonharm}, only even harmonics of $\Delta\omega$ occur in
$W(t)$ and in $\rho$.
This leads, according to Eq. (\ref{eq:13}), to
generation of new frequencies $\omega_0+(2j+1)\Delta\omega$ and
$2\omega_0+(2j+2)\Delta\omega$, $j=0,\pm1,\pm2,\ldots$, by Brunel mechanism, as shown in
\reffig{fig:nonharm}. In the tunnel non-commensurate case, the general
formula \refeq{eq:13} applies.

\section{Discussion and conclusion \label{sec:concl}}

In conclusion, we have considered generation of the so-called Brunel
harmonics in one- and two-color fields, comparing the situations when
the ionization is fast subcycle tunneling process depending only on
the instantaneous field (tunnel ionization regime valid for high
fields), and the case when it has ``memory'' and depends on the
cycle-average intensity (multiphoton regime valid for lower
fields). We used approach of direct modeling of the classical net
current produced by the ionized electrons and have shown that, at
least in this classical approximation, in the multiphoton ionization
the Brunel harmonics vanish in the multiphoton ionization regime.
This shows that Brunel harmonics are very
sensitive to the internal dynamics of the ionization
process. For the single-color fields, we were able to show the
transition between these two limits using Ivanov-Yudin approach which
gives correctly both ionization rate and ionization dynamics.
We confirmed that this behavior is also valid for two-color pump
pulses containing a fundamental frequency and its second
harmonics. Such  pump pulses are especially interesting
because they allow to generate THz frequencies, which are otherwise
difficult to access by other optical methods. In this case, we
considered only limiting cases of purely tunnel or purely multiphoton regime.

Nevertheless, for the two-color fields with the
incommensurate frequencies ($\omega_0$ and $\omega_0+\Delta\omega$)
and long enough pulses, the Brunel harmonics of the form
$\omega_0+j\Delta\omega$ (for integer $j$) can arise also in
multiphoton regime, because of the slow beatings between two harmonics
on the time scale $\sim1/\Delta\omega$.  In general, by changing of this beat
frequency, one can thus switch from ``typical tunnel'' to
``typical multiphoton'' regime, giving us a new tool to study the transition region
between the two ionization regimes without changing the
intensity which is an interesting direction for the future studies. 

\section*{Funding}

We gratefully acknowledge support by the DFG BA 4156/4-1, MO
850/20-1 and Nieders. Vorab ZN3061.

\appendix

\section{Two-color multiphoton ionization \label{sec:multiphot}} 

For description of multiphoton ionization we adopt here the Guo, Aberg and
Crasemann (GAC) model \cite{guo92}, which is the generalization of the
well known Keldysh-Faisal-Reis  theory
\cite{keldysh65,faisal73,reiss80} for the field of sufficiently
different frequencies and which was used in an optical setup \cite{tzankov07,babushkin08}. 
The GAC model is based on the  scattering-matrix formalism, 
providing non-perturbative treatment of 
electron in  electromagnetic field. In this section the units are used
in which $\hbar=1$, $c=1$. It provides the cycle-average
ionization rate for all intensities but fails to resolve the fast
subcycle ionization rate dynamics in the tunnel ionization regime. In our case, we have two-color
field  (for instance $\omega + 2\omega$) field. 
The Volkov state of the  electron in bichromatic laser field 
with a quantized vector potential $\vect A$: 
\begin{equation}
\label{eq:vect-potential}
\vect A(-\vect k \vect r) = \sum_{j=1}^2 g_i(\vec \epsilon_j e^{i
\vect k_j \vect r} a_j + \vec \epsilon_j^{\;\ast} e^{i
\vect k_j \vect r} a^\dag)
\end{equation}
with polarization $\vec \epsilon_j = [\vec \epsilon_x \cos(\xi_j/2)+i \vec \epsilon_y
\sin(\xi_j/2)] \exp(i \phi_j)$, defined by ellipticity $\xi_j$ and
phase $\phi_j$ for the modes $j=\omega,2\omega$,
can be written as \cite{guo92,gao98}
\begin{widetext}
\begin{equation}
\label{eq:volkov}
\left| \Psi_\mu \right> = V_e^{-1/2} \sum_{j_1,j_2} \exp\{i[\vect P_e +
(u_{p1}-j_1) \vect k_1 + (u_{p2}-j_2) \vect k_2]\} \mathcal{X}_{j_1,j_2}(z) \left| n_1
+ j_1, n_2 + j_2 \right>, 
\end{equation}
\end{widetext}
where $V_e$ is an integration volume, $n_i$ is a background photon number in the $i$th mode, $j_i$ is
the number of transferred photons, $\vect P_e$ us the momentum of the electron in the field, $\vect
k_1$, $\vect k_2$ are the wavevectors of the
corresponding field components, and 
$u_{p1}$, $u_{p2}$ are the ponderomotive parameters of
the electron in the fields $\omega$ and $2 \omega$ correspondingly: 
\begin{equation}
\label{eq:upi}
u_{pi} = \frac{e^2 \Lambda_i^2}{m_e \omega_i},
\end{equation}

where $i=1,2$, $\omega_1=\omega$, $\omega_2=2\omega$, $2\Lambda_i$
is an amplitude of the $i$th mode, obtained from a quantized field
\refeq{eq:vect-potential} in the large photon limit $g_i \sqrt{n_i} \rightarrow
\Lambda_i$. 

$\mathcal X_{j_1, j_2}(z)$ are the generalized phased Bessel functions (GPB) \cite{gao98,zhang04}
\begin{widetext}
  \begin{equation}
    \label{eq:GPB}
    \mathcal X_{j_1, j_2}(z) = \sum_{q_1\ldots q_4 = -\infty}^{\infty} \left\{ \mathrm
      X_{-j_1+2q_1+q_3+q_4}(\zeta_1) X_{-j_2+2q_2+q_3-q_4}(\zeta_2) \prod_{i=2}^4 \mathrm
      X_{-q_i}(z_i) \right\},
  \end{equation}
\end{widetext}
where $\mathrm X_n(z) = J_n(|z|)\exp\{i n \arg(z)\}$ is a phased Bessel
function \cite{zhang04}, calculated through the usual Bessel function
$J_n$ with  arguments
\begin{eqnarray}
\label{eq:GPBargs:zz} 
\, & \zeta_i   =  \frac{2 |e| \Lambda_i}{m_e \omega_i} \vect P \vec \epsilon_i, &
 z_i  =  \frac{1}{2}u_{pi} \vec \epsilon_i \vec \epsilon_i, \;\;\;\; (i=1,2)\\
\label{eq:GPBargs:z34}
\, & z_3  =  \frac{2 e^2 \Lambda_1 \Lambda_2 \vec \epsilon_1 \vect
\epsilon_2}{(\omega_1+\omega_2) m_e},  & 
z_4  =  \frac{2 e^2 \Lambda_1 \Lambda_2 \vec \epsilon_1 \vect
\epsilon_2^\ast}{(\omega_1-\omega_2) m_e}.
\end{eqnarray} 

Using the interaction Hamiltonian of the electron and the quantized
field $\vect A$:
\begin{equation}
\label{eq:hamilton}
V(\vect r) = \frac{-e}{2 m_e} \left[ (-i \vect \nabla)\vect A + \vect  A (-i
\vect \nabla ) \right] + \frac{e^2 \vect A^2}{2 m_e},
\end{equation}
one can obtain the transition amplitude from the ground state inside
an atom, 
described by a wavefunction $\Phi_i(\vect r)$ and the energy $\mathcal
E_i$, and  $l_1,\, l_2$
photons in the corresponding modes of the field, to an electron free state with a
wavefunction $\phi_f(\vect r) = e^{i \vect P_{ef} \vect r}$ and the
energy $\mathcal E_f$, and $m_1,
\, m_2$   photons  in the field  \cite{guo92}. 
\begin{widetext}
  \begin{equation}
    \label{eq:trans} T_{fi} =   V_e^{-1/2} \sum_{\varepsilon_\mu=\varepsilon_i}
    \left<\phi_f(\vect r); m_1,m_2 | \Psi_\mu \right> \left< \Psi_\mu
      | V(\vect r) | 
      \Phi_i(\vect r); l_1,l_2 \right> =  V_e^{-1/2} \mathcal B_{q_1,q_2}^{(\omega_1,\omega_2)}(\vect P_{ef}),
  \end{equation}
\end{widetext}
where $q_i = l_i-m_i$, and   $\mathcal B_{q_1,q_2}^{(\omega_1,\omega_2)}(\vect P_{ef})$ is an (unnormalized) electron transition
amplitude to the final state $\phi_f(\vect r)$, accomplished by
absorption of exactly $q_i$ photons from the $i$-th field mode:
\begin{widetext}
  \begin{equation}
    \label{eq:b}
    \mathcal B_{q_1,q_2}^{(\omega_1,\omega_2)}(\vect P_{ef}) =  \Phi_i(\vect P_{ef}) \sum_{j_1,j_2}
    \left[ (u_{p1}-j_1) \omega_1 + (u_{p2}-j_2) \omega_2
    \right] \mathcal X_{j_1-q_1,j_2-q_2}(z_f) \mathcal X^\ast_{j_1,j_2}(z_f),
  \end{equation}
\end{widetext}
with $\Phi_i(\vect P_{ef})$ being  a Fourier transform of $\Phi_i(\vect
r)$. 

The total transition rate is then given by integration over all
possible electron momenta and is expressed as a sum over the partial
probabilities of absorption of particular number of ``biphotons''
$q=q_1+2q_2$ leading finally to: 
\begin{eqnarray}
\nonumber 
&& W_{GAC} = \frac{V_e}{(2 \pi)^2} \int \int d\Omega \: d|\vect P_{ef}|\: \left\{ |\vect P_{ef}|^2 \delta(\mathcal E_i
-\mathcal E_f) | T_{fi}|^2 \right\} =  \\
&& \frac{\sqrt{2 m_e^3 \omega}}{4 \pi^2} \sum_q 
\sqrt{q-B/\omega} \int d\Omega  \left|
\sum_{q_2} \mathcal B_{q-2 q_2,q_2}^{(\omega,2 \omega)}(\vect p_{ef})
\right|^2. 
\label{eq:W}
\end{eqnarray}
Here,
$B$ is an ionization potential, $\Omega$ is a solid angle ($d\Omega = \sin \theta_{\vect p_{ef}}
d\theta_{\vect p_{ef}} d \phi_{\vect p_{ef}}$), and
$\vect p_{ef} =  \vect n_\Omega \sqrt{2 m_e \omega (q-B/\omega)}$,
$\vect n_\Omega$ is a unit vector, pointing in the direction defined
by the angle $\Omega$.


%

\end{document}